\documentclass[%
 reprint,
superscriptaddress,
 amsmath,amssymb,
 aps,
]{revtex4-1}

\usepackage{graphicx}
\usepackage{dcolumn}
\usepackage{bm}

\usepackage[T1]{fontenc}
\usepackage[colorlinks,bookmarks=false,citecolor=blue,linkcolor=red,urlcolor=blue]{hyperref}
\usepackage[dvipsnames]{xcolor}
\usepackage{times}
\usepackage{bbm}

\newcommand{\be}{\begin{eqnarray}}
\newcommand{\ee}{\end{eqnarray}}

\begin{document}


\title{Competing Electronic Phases near the Onset of Superconductivity in Hole-doped SrFe$_2$As$_2$}

\author{L.~Wang}
\affiliation{Institute for Solid State Physics, Karlsruhe Institute of Technology, 76021 Karlsruhe, Germany}%
\affiliation{Kirchhoff Institute of Physics, Heidelberg University, INF 227, 69120 Heidelberg, Germany}

\author{M.~He}
\affiliation{Institute for Solid State Physics, Karlsruhe Institute of Technology, 76021 Karlsruhe, Germany}%
\affiliation{Chongqing Key Laboratory of Soft Condensed Matter Physics and Smart Materials, College of Physics, Chongqing University,Chongqing 401331, People's Republic of China}%

\author{D.~D.~Scherer}
\affiliation{Niels Bohr Institute, University of Copenhagen, Lyngbyvej 2, DK-2100 Copenhagen, Denmark}%

\author{F.~Hardy}
\author{P.~Schweiss}
\author{T.~Wolf}
\author{M.~Merz}
\affiliation{Institute for Solid State Physics, Karlsruhe Institute of Technology, 76021 Karlsruhe, Germany}

\author{B.~M.~Andersen}
\affiliation{Niels Bohr Institute, University of Copenhagen, Lyngbyvej 2, DK-2100 Copenhagen, Denmark}

\author{C.~Meingast}
 \email{christoph.meingast@kit.edu}
\affiliation{Institute for Solid State Physics, Karlsruhe Institute of Technology, 76021 Karlsruhe, Germany}


\date{\today}

\begin{abstract}
An intriguingly complex phase diagram of Na-doped SrFe$_2$As$_2$ is uncovered using high-resolution thermal-expansion, magnetization and heat-capacity measurements. The detailed temperature dependence of the orthorhombic distortion and the anisotropy of the uniform magnetic susceptibility provide evidence for nine distinct electronic phases near the transition region between stripe antiferromagnetism and unconventional superconductivity. In particular, we report the finding of a new magnetic phase which competes surprisingly strongly with superconductivity. From theoretical studies we propose that this phase is a double-${\bf Q}$ phase consisting of a mixture of symmetry-distinct commensurate magnetic orders with a peculiar temperature-dependent magnetic moment reorientation.
\end{abstract}

\pacs{Valid PACS appear here}
\maketitle



The fascinating interplay between magnetism and superconductivity, and the possible role of magnetic fluctuations in driving unconventional superconductivity, remain research topics of considerable interest. Iron-based superconductors (FeSCs) in particular constitute an important class of materials where many open questions related to magnetism and superconductivity still remain ~\cite{dai}. In addition to the outstanding question of the underlying mechanism of superconductivity, the origin of electronic nematicity and the detailed nature, and proper theoretical description, of magnetism in these materials remain examples of current controversy~\cite{RN143}. 

A promising gateway to understand the dominant low-energy electronic interactions in FeSCs is found through a study of the plethora of novel magnetic phases recently discovered in these materials~\cite{avci14a,bohmer15a,allred15a,meier17,wasser15,Sr122Thomas2014}. This line of progress was exemplified in the theoretical prediction~\cite{lorenzana08,eremin}, and subsequent experimental discovery, of the so-called $C_4$-symmetric magnetic phases in K- and Na-doped BaFe$_2$As$_2$, SrFe$_2$As$_2$, and more recently also CaFe$_2$As$_2$ ~\cite{TaddeiPRB2017}. The prevalent magnetic phase of FeSCs is a metallic single-${\bf Q}$ spin-density wave (SDW) with ordering wave vectors ${\bf Q}_{1}= (\pi,0)$ or ${\bf Q}_{2} = (0,\pi)$, also known as the magnetic stripe (MS) phase. This collinear MS phase breaks $C_4$ tetragonal symmetry and therefore exists in an orthorhombic Fe crystal structure, with the in-plane magnetic moments oriented ferromagnetically (antiferromagnetically) along the shorter (longer) Fe-Fe bonds~\cite{dai}. The $C_4$-symmetric double-${\bf Q}$  phases consist of magnetism condensed simultaneously at both ordering wave vectors ${\bf Q}_{1}$ and ${\bf Q}_{2}$, resulting in either 1) a collinear charge- and spin- ordered density wave (CSDW) phase, or 2) a non-collinear spin-vortex crystal (SVC) phase with magnetic moments on neighboring sites at right angles to each other.

While the three distinct commensurate (C) magnetic phases mentioned above seem by now reasonably well understood within itinerant electron models~\cite{lorenzana08,eremin,brydon2011,giovannetti,gastiasoro15,scherer16,hoyer,christensen17,gastiasoro17,halloran,allred16a}, the focus has shifted to outstanding questions related to newly discovered magnetic phases. For example, in Na-doped BaFe$_2$As$_2$ a magnetic phase dubbed $C_2'$ was identified in a mosaic of other phases near the foot of the magnetic dome, by high-resolution thermal-expansion and specific-heat measurements~\cite{LiranPRB2016}. This $C_2'$ phase motivated a recent theoretical study of the possible incommensurate (IC) magnetic structures allowed in FeSCs, resulting in the proposal of several entirely new textured magnetic phases possibly realized in these materials~\cite{christensen18}. Thus, it is important to hunt down the nature of new novel phases in FeSCs, both for the general understanding of these materials but also for finding a fitting  theoretical description of magnetism in FeSCs. Important open questions in this regard include the following: 1) are the novel magnetic phases discovered in  Na-doped BaFe$_2$As$_2$ ubiquitous to other hole-dped FeSCs, or rather a peculiarity of this specific material?, 2) do new phases (e.g. mixed or vestigial)  exist well below the magnetic critical transition temperature?, 3) what is the nature of the transition between different magnetic phases?, and 4) what is the interplay between superconductivity and double-${\bf Q}$ magnetism? 

Here, we address these open questions through a detailed mapping of the phase diagram of Na-doped SrFe$_2$As$_2$ combining high-quality single crystals, thermodynamic probes as well as theory. We find that indeed the phases previously discovered in Na-doped BaFe$_2$As$_2$ are also found in Na-doped SrFe$_2$As$_2$, however our measurements additionally demonstrate that the phase diagram of Na-doped SrFe$_2$As$_2$ exhibits even more complexity than the K- and Na-doped BaFe$_2$As$_2$ counterparts~\cite{bohmer15a,LiranPRB2016}. Specifically, we identify the emergence of a new magnetic phase in Na-doped SrFe$_2$As$_2$.  We interpret this state as a mixed double-${\bf Q}$ magnetic phase and use theoretical modeling to propose a temperature evolution of the magnetic moments that is consistent with both the thermal expansion and uniform susceptibility measurements.

\begin{figure*}[]
\centering
\begin{minipage}{2\columnwidth}
\centering
\includegraphics[width=1\columnwidth]{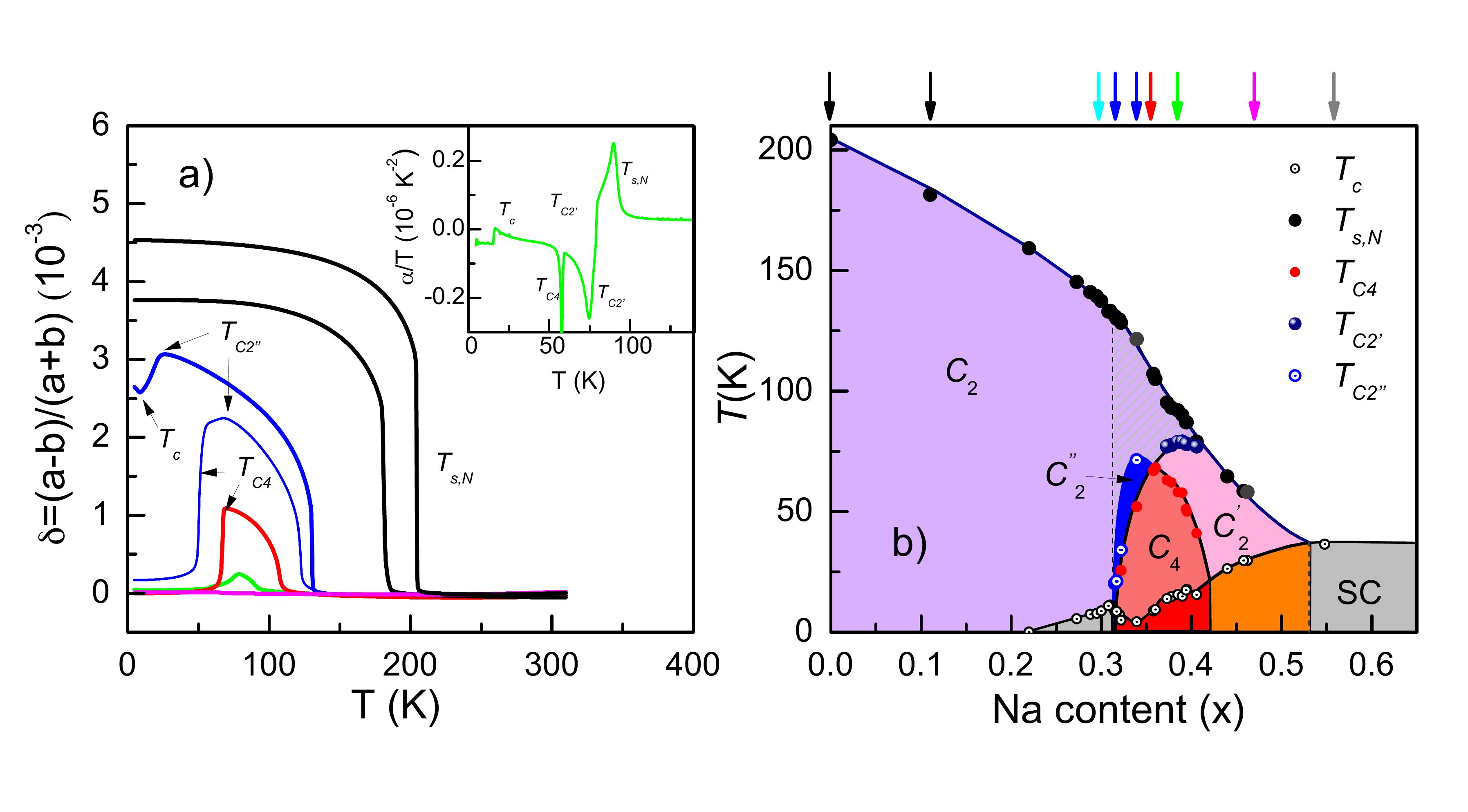}
\end{minipage}
\caption{(a) Orthorhombic distortion, $\delta =(a-b)/(a+b)$, derived from dilatometry (for x = 0, 0.11, 0.32, 0.34, 0.36, 0.39, and 0.46) and (b) resulting phase diagram from thermal expansion, specific heat and magnetization measurements. The concentrations of the crystals shown in a) are marked by the color-coded arrows in b) and matching colors will be used throughout the paper.}
\label{fig:PD}
\end{figure*}

\begin{figure}[t]
\centering
\begin{minipage}{1\columnwidth}
\centering
\includegraphics[width=1\columnwidth]{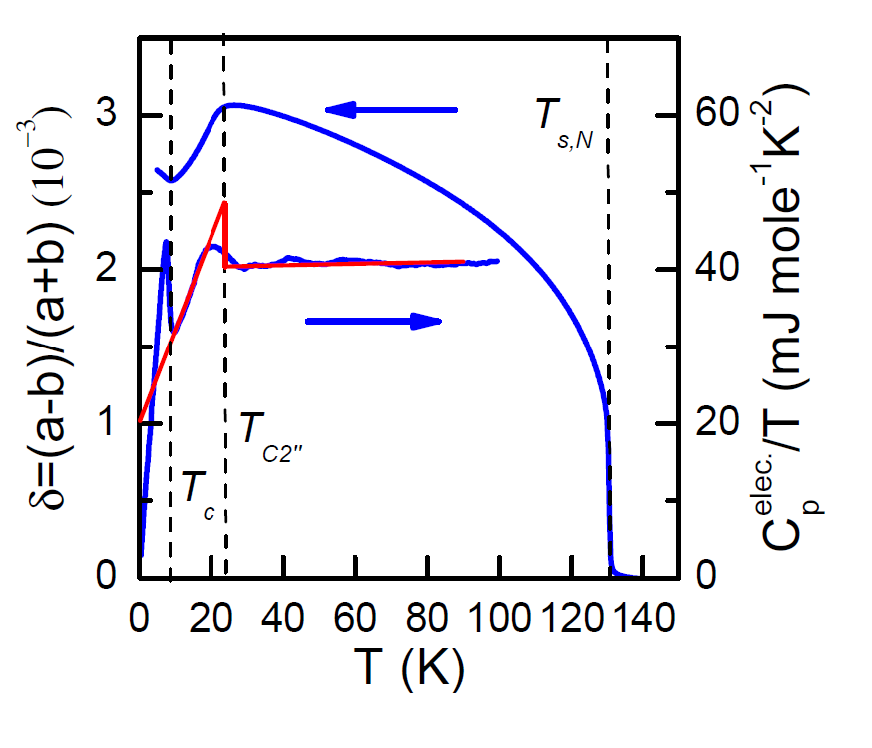}
\end{minipage}
\caption{Orthorhombic distortion, $\delta$, and electronic heat capacity, $C_e/T$, for the crystal with $x = 0.32$  showing the new magnetic phase at $T$ = 23 K.  This phase is characterized by a reduction of $\delta$ and a significant suppression of electronic density of states, as indicated by the red lines.  Superconductivity sets in at 9.6 K and results in an increase of $\delta$.}
\label{fig:Cp}
\end{figure}

\begin{figure}[t]
\centering
\begin{minipage}{1\columnwidth}
\centering
\includegraphics[width=0.8\columnwidth]{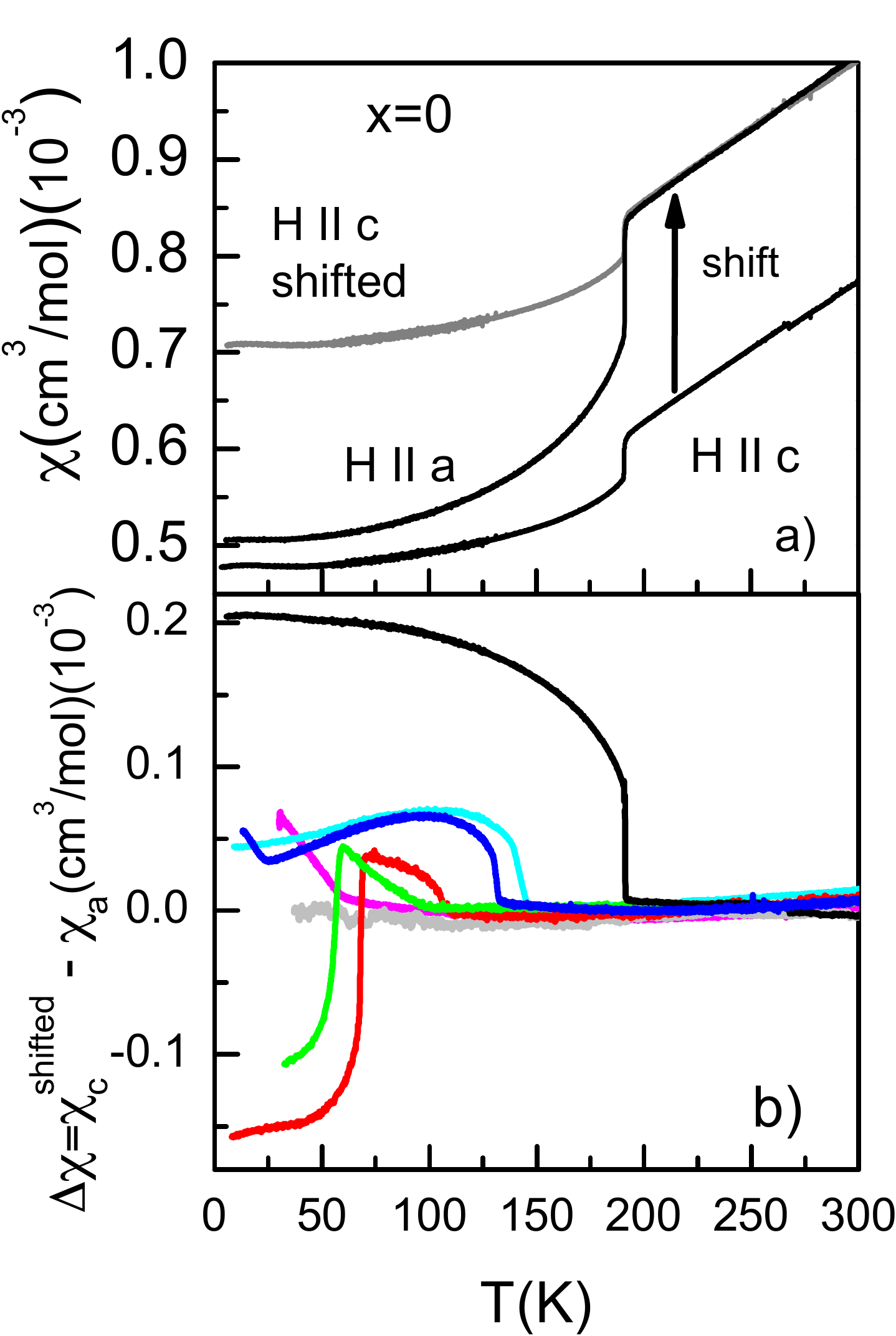}
\end{minipage}
\caption{(a) Uniform magnetic susceptibility  for $x = 0$ in a field of 12 T applied parallel to the $a$ and $c$ axes.  (b) The difference in susceptibilites, $\Delta \chi = \chi^{c} - \chi^{a} $  ( $ \chi^{c} $  has been shifted vertically to match $ \chi^{a} $ above the SDW transition) for various Na-contents (see Fig. 1 for color coding).  $\Delta \chi$ is a sensitive measure of magnetic order.  (see text for details).}
\label{fig:Magnetization}
\end{figure}

Single crystals of Na-doped SrFe$_2$As$_2$ were grown using a self-flux method~\cite{Hardy2010,Sr122Thomas2014} and characterized using capacitance dilatometry, heat capacity, magnetization, and single-crystal x-ray diffraction, as described in more detail in the supplemental section~\cite{SM}.	In Fig.~\ref{fig:PD}a the orthorhombic distortion, $\delta = (b-a)/(b+a)$, which is a very sensitive indicator of the structural response to magnetic order, derived from our thermal expansion data~\cite{bohmer15a,LiranPRB2016} is plotted versus temperature for various doping levels. The reduction of $T_{\mathrm{N}}$ with Na-doping and the emergence of the double-${\bf Q}$ $C_4$ reentrant phase, in which the orthorhombicity suddenly vanishes (see e.g. the red curve), are both similar to what has been observed previously in Ba$_{1-x}$Na$_x$Fe$_2$As$_2$\cite{LiranPRB2016}.  The inset in Fig.~\ref{fig:PD}a, in which we show the thermal expansivity curve for $x = 0.32$ exhibiting four distinct phase transitions, demonstrates both the high quality of the crystals, as indicated by the sharp transitions, as well as the high sensitivity of our thermal-expansion setup.  The phase diagram constructed from our thermodynamic measurements (Fig. \ref{fig:PD}b) is at first glance remarkably similar to that of Ba$_{1-x}$Na$_x$Fe$_2$As$_2$~\cite{LiranPRB2016}. In particular for $x > 0.36$, we also find clear evidence for the yet unidentified $C_2'$ phase, as well as the inflection in the slope of $T_{\mathrm{N}}$ versus $x$ near the onset of the additional phases.  However, we could also resolve a small region on the left side of the $C_4$ reentrant phase which contains a new magnetic phase, which to our knowledge has not been observed previously. This phase, which we label $C_2''$, emerges out of the stripe phase via a second-order transition at 25 K and has a reduced orthorhombicity (see blue curves in Fig.~\ref{fig:Magnetization}a and Fig.~\ref{fig:Cp}). 

The plot of $C_e/T$ vs $T$ in Figure~\ref{fig:Cp} demonstrates that there is a considerable loss of electronic density of states associated with the $C_2''$ transition.  The superconducting transition in the heat capacity occurs at 8.5 K  and results in an increase of the orthorhombic distortion below $T_c$, which is an indication for a strong competition between superconductivity and the $C_2''$ phase. In the following we will concentrate on this new phase and a more detailed study covering the data of the whole phase diagram will be presented separately~\cite{WangUnpub}.

In order to uncover the nature of the $C_2''$ phase, we performed measurements of the uniform magnetization for magnetic fields both parallel and perpendicular to the crystallographic $c$-axis (see Fig.~\ref{fig:Magnetization}).  The magnetization at 12 Tesla above the SDW transition has exactly the same temperature dependent slope for both field directions and for all values of $x$ (see Fig.~\ref{fig:Magnetization}a and supplemental), but are presumably shifted relative to one another by an anisotropic temperature-independent Van-Vleck term.  By taking the difference in susceptibilities for fields parallel and perpendicular to $c$ (after a shift to eliminate the Van-Vleck term), a very sensitive indicator of the magnetic ordering is obtained, as shown in Fig.~\ref{fig:Magnetization}b for several compositions.  As shown previously, a sizable uniform magnetic susceptibility anisotropy in an itinerant system can only be expected in a magnetically ordered state and provides information about the spin orientation~\cite{meingast1,hou17} .

  A larger susceptibility is expected for fields perpendicular to the spin orientation, in agreement with the data for undoped SrFe$_2$As$_2$, for which $\tilde{\chi}_c$ > $\tilde{\chi}_a$ implying in-plane spin orientation (see Fig.~\ref{fig:Magnetization}b)\cite{HouJ2017}. On the other hand, the spin orientation is known to flip to the $c$-axis in the double-${\bf Q}$ reentrant $C_4$ phase~\cite{wasser15,malletta,mallettb}, which is consistent with the sign change in $\tilde{\chi}_c$-$\tilde{\chi}_a$ observed for the red and green curves in Fig.~\ref{fig:Magnetization}b.  The susceptibility anisotropy of the new phase (dark blue line in Fig.~\ref{fig:Magnetization}b) has an interesting temperature dependence, first increasing at $T_{\mathrm{N}}$, then going over a maximum and slightly decreasing and finally increasing again below $T_{C2''}$. The always positive values indicate mostly in-plane magnetic order, and importantly, the increase below $T_{C2''}$ clearly demonstrates that this phase is distinct from the usual double-${\bf Q}$ $C_4$ phase. Finally, the susceptibility difference is also positive for the $C_2'$ phase, proving that this phase is 1) magnetic and 2) has in-plane order, which suggests that it may be a derivative of a spin-vortex type ordering\cite{meier17}.

%



In order to gain insight into the temperature evolution of the magnetic
state at the doping level $x = 0.32$, corresponding to the blue scan in Fig.~\ref{fig:Magnetization}b,
we use a simple phenomenological model for itinerant magnetic states. For all the details of the theoretical calculation, 
we refer to the SM section~\onlinecite{SM}. The Hamiltonian of the model contains the kinetic part for the Fe $3d$ electrons~\cite{ikeda10} (with orbital character of $xz$, $yz$, $x^2-y^2$, $xy$, $3r^2-z^2$ symmetry with respect to the Fe square lattice), and a mean-field self-energy that describes the details of the coupling of electrons to the SDW magnetization with ordering vectors ${\bf Q}_{1}$ and ${\bf Q}_{2}$, respectively. Here, we focus exclusively on the hole-doped regime of the model. The Fourier components of the SDW magnetization can be interpreted as magnetic moments 
${\boldsymbol m}_{1}$ and ${\boldsymbol m}_{2}$, that measure the strength and orientation of the magnetic order relative to the lattice. Within the itinerant
model, a finite SDW order leads to a reconstruction of the electronic band structure,
which ultimately leaves signatures in key observables of the electronic system.
For a given magnetic configuration, we compute the magnetic susceptibility in a simple
approximation as the summed particle-hole propagator for the mean-field dressed electron Greens function. We then show that with a suitable choice for the temperature evolution of the magnetic moments and their orientation, we can capture the temperature dependence of the measured magnetic anisotropy at a qualitative level.

\begin{figure}[t]
\centering
\begin{minipage}{1\columnwidth}
\centering
\includegraphics[width=1\columnwidth]{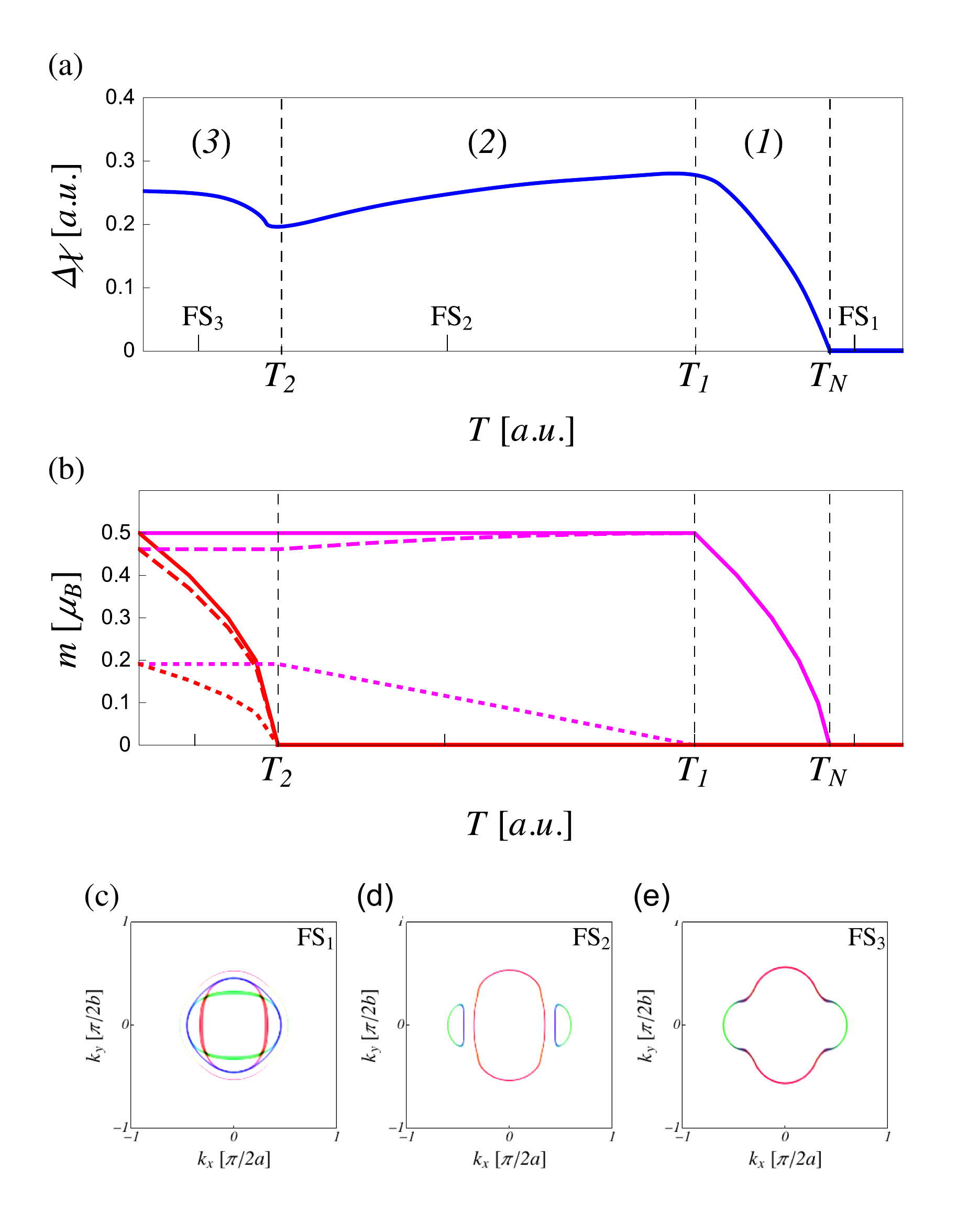}
\end{minipage}
\caption{(a) Evolution of the magnetic out-of-plane anisotropy as a function
of temperature, as determined from the phenomenological model. (b) Assumed evolution of magnetic moments associated with SDW order, where magenta and red solid curves 
show $ | {\boldsymbol m}_{1} | $ and  $ | {\boldsymbol m}_{2} | $, respectively. Dashed and dotted curves of the corresponding color show the moments' $x$ and $z$ components. (c)-(e) Evolution of Fermi surfaces FS$_{1}$-FS$_{3}$ in the reduced Brillouin zone of the phenomenological model under the influence of band reconstruction due to SDW order. The Fermi surfaces are determined from the orbitally resolved spectral weight. Red and green colors correspond to dominating $xz$ and $yz$ orbital weight, respectively. The blue Fermi surfaces segments highlight regions with dominating $xy$ orbital character.}
\label{fig:theory}
\end{figure}

In modelling the magnetic anisotropy, we assume three distinct regimes, corresponding
to different types of changes in the underlying magnetic state, in order to describe
the main features observed in the anisotropy versus temperature, as shown in Fig.~\ref{fig:theory}.

\noindent \textit{Regime 1}: Upon entering the magnetic state below $T_{\mathrm{N}}$, we assume a $C_{2}$-symmetric single-${\bf Q}$ state with ${\bf Q}_{1}$ ordering vector and the moment parallel to the $a$ direction of the Fe square lattice. As we choose to work with a coordinate system, where $x = a$, $y = b$ and $z = c$, we write
${\boldsymbol m}_{1} \parallel \hat{{\boldsymbol e}}_{x} $. Starting from a 
vanishing moment at $T_{\mathrm{N}}$, we then assume an increasing moment
$  | {\boldsymbol m}_{1} | \sim  | T - T_{\mathrm{N}}  |^{1/2} $
as the temperature $T$ is lowered, reaching its saturation value at temperature $T_{1}$.
Given that hole-doping typically leads to a reduction of the magnetic moment
compared to the parent compound, we assume that saturation occurs
at a value of $ m_{0} = 0.5 \mu_{\mathrm{B}} $. Note that the growing $C_4$-symmetry breaking in this regime leads to an enhanced orthorhombic distortion, as seen experimentally in Fig.~\ref{fig:PD}a.

\noindent \textit{Regime 2}: As the system is cooled down to a temperature
below $ T_ {1} $, we assume that -- by some microscopic mechanism (e.g. spin-orbit coupling~\cite{christensen15,scherer2018,borisenko}) not
 explicitly included in our phenomenological model -- the magnetic moment starts tilting
out of the $ab$ plane and develops a finite projection along $c$. Denoting
the tilt angle by $\theta$, we have $ {\boldsymbol m}_{1} = m_{0} 
(\cos(\theta) \hat{{\boldsymbol e}}_{x} + \sin(\theta) \hat{{\boldsymbol e}}_{z} )$.
The tilting angle then continuously increases with decreasing temperature. While we cannot easily pinpoint the final value of the tilting angle, it must satisfy $ \theta < \theta^{\ast} $ for an angle $ \theta^{\ast} < \pi/2 $, which will be discussed below. We assume, that the out-of-plane tilting eventually stops at a temperature $ T_{2} < T_{1} $ at an angle $\theta_{0}$. The further enhancement of the orthorhombic distortion seen experimentally in this regime, can be captured by a slightly growing overall moment as $T$ is lowered. For simplicity we have not included this in the model since it is unimportant for the magnetic anisotropy.

\noindent \textit{Regime 3}: With the temperature decreasing below $T_{2}$, 
the moment of the second SDW component, $ {\boldsymbol m}_{2} $, is assumed to 
grow continuously. For simplicity, we restrict the analysis to collinear configurations,
$ {\boldsymbol m}_{1} \parallel {\boldsymbol m}_{2}  $. We let the second SDW component saturate at $m_{0}$ as well, such that the low-temperature magnetic state is characterized by moments $ {\boldsymbol m}_{1} = {\boldsymbol m}_{2} = m_{0} 
(\cos(\theta_{0}) \hat{{\boldsymbol e}}_{x} + \sin(\theta_{0}) \hat{{\boldsymbol e}}_{z} ) $ and fully restored $C_{4}$ symmetry. The reduction of the $C_4$ symmetry breaking in this $T$-regime, naturally explains the reduced orthorhombic distortion at $T<T_2$ ($T_{C2''}$ in Fig. 1 and Fig.2).

In Fig.~\ref{fig:theory}(a) we show the resulting temperature dependence
of the out-of-plane anisotropy for our phenomenological model, here defined as
\be 
\Delta \chi = \Delta\chi^{zz} - \frac{1}{2} \left( \Delta\chi^{xx} + \Delta\chi^{yy} \right).
\ee
The temperature dependence of the magnetic moments is shown in Fig.~\ref{fig:theory}(b).
In regime 1, we observe the increase of $\Delta\chi$ as the SDW order develops.
The saturation of the moment $ {\boldsymbol m}_{1} $ coincides with the
maximum of $\Delta\chi$ at $T_{1}$. The subsequent tilting of the moment in regime 2 leads to a reduction of $\Delta\chi$, followed by a small increase upon the onset of the second SDW component below $T_{2}$ in regime 3. Our model calculation gives access to the band reconstruction of the Fermi surface. In Fig.~\ref{fig:theory}(c)-(f), we show the evolution of the Fermi surface in the paramagnetic state over the $C_{2}$ MS state to an almost $C_{4}$ symmetric CSDW state.

Above, we have only shown results for a single chemical potential realizing a hole-doped situation by a rigid band shift. We performed analogous simulations for different values of the chemical potential, while keeping the magnetic configurations fixed. The results shown above appear to be stable for a certain window of chemical potentials, but both on the hole- and electron-doped sides, the resulting out-of-plane anisotropy is eventually modified rather  drastically and can even undergo a complete sign reversal for all temperatures. While we have not performed a self-consistent calculation (and the favored magnetic order certainly responds to changes in the doping level), these results point to the sensitivity of the anisotropy to Fermi surface details. We have additionally performed simulations for larger moments with $m_{0} = 1 \mu_{\mathrm{B}}$, but found qualitatively similar results as for the case $m_{0} = 0.5 \mu_{\mathrm{B}}$.  

Above, we introduced the angle $\theta^{\ast}$, where, starting from $\Delta\chi > 0$,
one finds $\Delta\chi = 0$ as $\theta$ approaches $\theta^{\ast}$ from below. Upon further increasing the tilting angle, the anisotropy turns negative, $\Delta\chi < 0$. The precise value of $\theta^{\ast}$ depends on microscopic details. We note, that the switching in the sign of the out-of-plane anisotropy is related to rotational symmetry about the $y$-axis (as long as SOC effects are negligible). But the value of $\theta^{\ast}$ is determined by the actual value of $\Delta\chi^{yy}$. For the moment oriented along $x$, we find $\Delta\chi^{zz} = \Delta\chi^{yy} > 0$, $\Delta\chi^{xx} < 0 $ together with $  | \Delta\chi^{xx} |  =  | \Delta\chi^{zz} | $. The out-of-plane tilting then corresponds to a rotation about the $y$-axis. This leaves $\chi^{yy}$ unaffected, but $\chi^{zz}$ decreases while $\chi^{xx}$ increases, until eventually $\chi^{xx} = \chi^{zz}$, implying $\Delta\chi^{xx} = \Delta\chi^{zz}$ = 0, since $ | \Delta\chi^{xx} |  =  | \Delta\chi^{zz} | $ still holds. But already at an angle, where $\Delta\chi^{zz} = 2/3 \Delta\chi^{yy}$, $\Delta\chi$ vanishes, leading to $\Delta\chi < 0$ upon further increase of the tilting angle.

This discussion can easily be extended to explain the qualitative behavior of the red and green temperature scans in Fig. \ref{fig:Magnetization}b, where the magnetic state shows a change in orientation from moments aligned along $x$, to moments fully aligned along $z$. First, we note that the arguments above hold, irrespective of whether we have single-${\bf Q}$ or collinear double-${\bf Q}$ magnetic order. Then, upon going from $x$-axis ($\theta = 0$) to $z$-axis ($\theta = \pi/2$) polarization, the anisotropic susceptibility components satisfy $\Delta\chi^{xx}(\pi) = \Delta\chi^{zz}(0)$ and $\Delta\chi^{zz}(\pi) = \Delta\chi^{xx}(0)$ along with $  | \Delta\chi^{xx} |  =  | \Delta\chi^{yy} | =  | \Delta\chi^{zz} | \equiv \gamma $.
Therefore, the out-of-plane susceptibility changes from $\Delta\chi(0) = \gamma $ to $ \Delta\chi(\pi) = -2\gamma$, in nice agreement with the jump observed in the experimental data.


In conclusion, through a detailed study of the phase diagram of (Sr,Na)Fe$_2$As$_2$ using thermodynamic probes we have uncovered an additional magnetic phase in Fe-based materials, distinct from the well-known stripe or double-${\bf Q}$ $C_4$ phases.  We suggest that this is a mixed double-${\bf Q}$ phase with some degree of out-of-plane tilting of the magnetic moments.  This is supported by theoretical calculations of the uniform susceptibility.  Similar to the other double-${\bf Q}$ phases, superconductivity is strongly suppressed in this phase, which we show is a direct consequence of a significantly reduced electronic density of states.  Similar to Na-doped BaFe$_2$As$_2$, we also observe the $C_2'$ phase in Na-doped SrFe$_2$As$_2$ and provide evidence of in-plane magnetic order in this phase.  Most likely, this phase is closely related to the spin-vortex type phase, which was recently discovered in the '1144' compound\cite{meier17}.  The fact that all these different magnetic phases naturally emerge out of a weak-coupling itinerant approach strongly suggests that the physics of superconductivity in Fe-based materials should also be accessible using this same approach.

\begin{acknowledgments}
We acknowledge discussions with Morten H. Christensen, Panagiotis Kotetes. The contribution from M.M. was supported by the Karlsruhe Nano Micro Facility (KNMF). L.W thanks the support from Deutsche Forschungsgemeinschaft(DFG) through Grant No.WA4313/1-1. D.D.S. and B.M.A acknowledge financial support from the Carlsberg Foundation. 
\end{acknowledgments}

\bibliographystyle{plain} 




\begin{widetext}

\newpage

\setcounter{equation}{0}
\setcounter{figure}{0}
\setcounter{table}{0}
\setcounter{page}{1}
\makeatletter
\renewcommand{\theequation}{S\arabic{equation}}
\renewcommand{\thefigure}{S\arabic{figure}}
\renewcommand{\bibnumfmt}[1]{[S#1]}
\renewcommand{\citenumfont}[1]{S#1}

\begin{center}
\textbf{\large Supplementary Material: ``Competing Electronic Phases near the Onset of Superconductivity in Hole-doped SrFe$_2$As$_2$''}
\end{center}

\section{S1. Experimental details}
\label{sm:model}

Single crystals of Na-doped SrFe$_2$As$_2$  were grown in alumina crucibles using a self-flux method with (Sr,Na): FeAs ratios 1:3.5--1:5. The crucibles were sealed in iron cylinders filled with argon gas. After heating to 1100 - 1180 $^\circ$C, the furnace was cooled down slowly at rates between 0.3 and 0.5 $^\circ$C/h to minimize the amount of flux inclusions. Near 920--1080 $^\circ$C the furnace was turned upside down to separate the remaining liquid flux from the grown crystals and then cooled down to room temperature with intermediate holds to in-situ anneal the crystals. 

Thermal expansion was measured using a high-resolution home-made capacitance dilatometer \cite{Meingast1990}, which is several orders of magnitude more sensitive than traditional diffraction techniques.

Heat capacity was measured using a Physical Property Measurement System (PPMS) from Quantum Design. The electronic specific heat was obtained by subtracting an appropriate phonon background \cite{Bohmer2015ab,Hardy2010,Hardy2013}.
 
The Na content of many crystals used for the thermal-expansion and specific-heat measurements was accurately determined by x-ray diffraction using a Stoe imaging plate diffraction system (IPDS-2T) equipped with Mo K-alpha radiation. All accessible symmetry-equivalent reflections ($\sim$1730) were measured at RT up to a maximum angle 2$\theta$ = 65$^{\circ}$. The data were corrected for Lorentz, polarization, extinction, and absorption effects. Using SHELXL\cite{Sheldrick2008} and JANA2006\cite{Pectric2014} around 100 averaged symmetry-independent reflections (I > 2$\sigma$) have been included for the respective refinements in space group I4/mmm. The refinements converged quite well and show excellent weighted reliability factors (wR-2) which are typically around 4.5. The Na content of the other crystals were interpolated between these fixed points using the SDW transition temperature as a reference. The values of the structural parameters from our x-ray refinement are in good agreement with previous results \cite{Avci2013}.

Magnetization measurements were made using the VSM option of the PPMS system.  The samples were attached to the sample holder using a small amount of grease, the amount of which was determined by weighing the difference of the sample holder with and without grease.  The magnetization signal arising from the grease, which was determined separately, was subtracted from each data set.  

\begin{figure*}[h!]
\centering
\begin{minipage}{1\columnwidth}
\centering
\includegraphics[width=0.82\columnwidth]{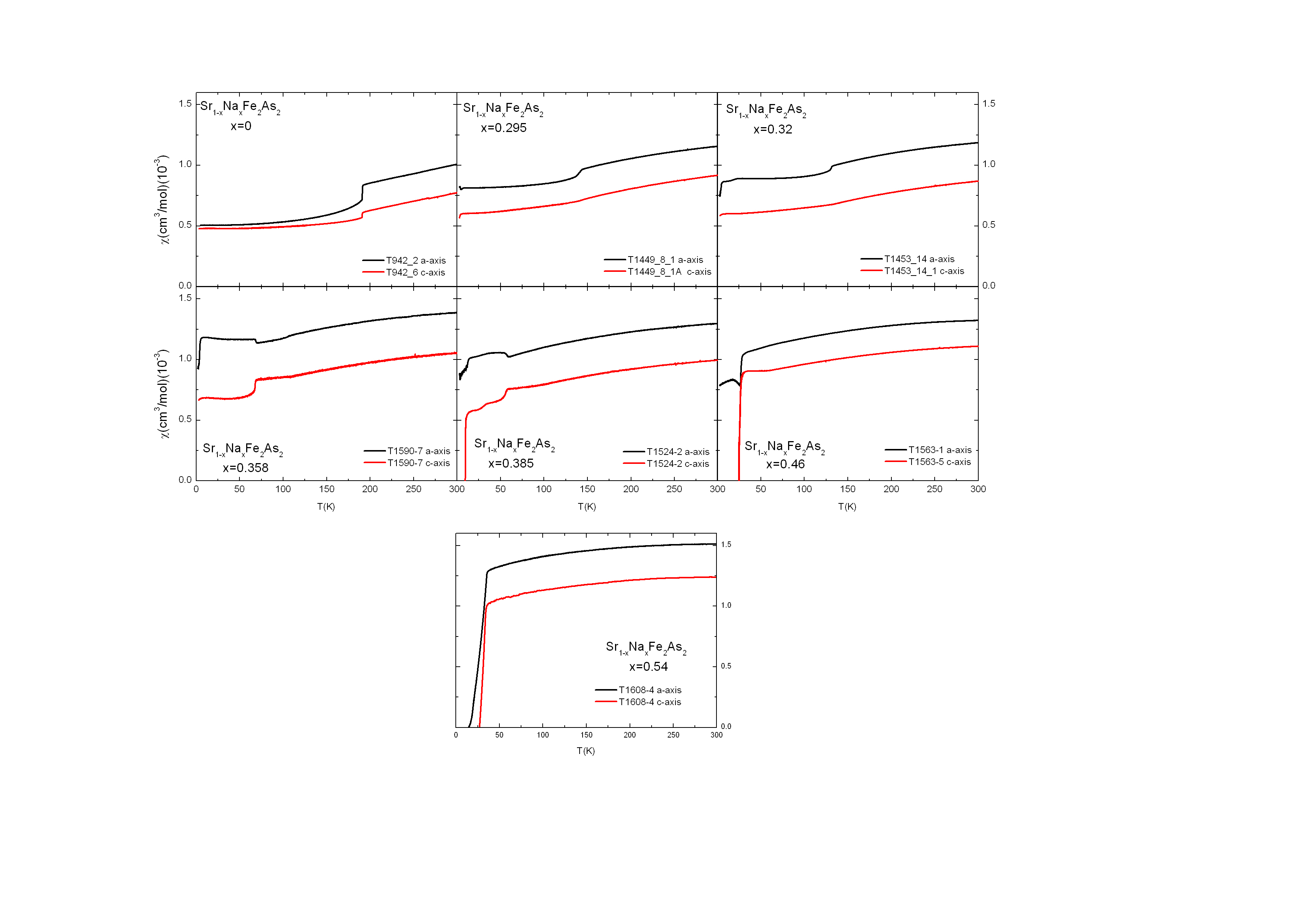}
\end{minipage}
\caption{Uniform magnetic susceptibility at 12 Tesla for fields parallel and perpendicular to the crystallographic $c$-axis after subtraction of the grease magnetization for several Na - concentrations.  Note that the difference between the two field directions above the SDW transition is nearly both temperature and doping independent.  We attribute this difference to a Van-Vleck term.}
\label{fig:experimental_1}
\end{figure*}
\begin{figure*}[h!]
\centering
\begin{minipage}{1\columnwidth}
\centering
\includegraphics[width=0.82\columnwidth]{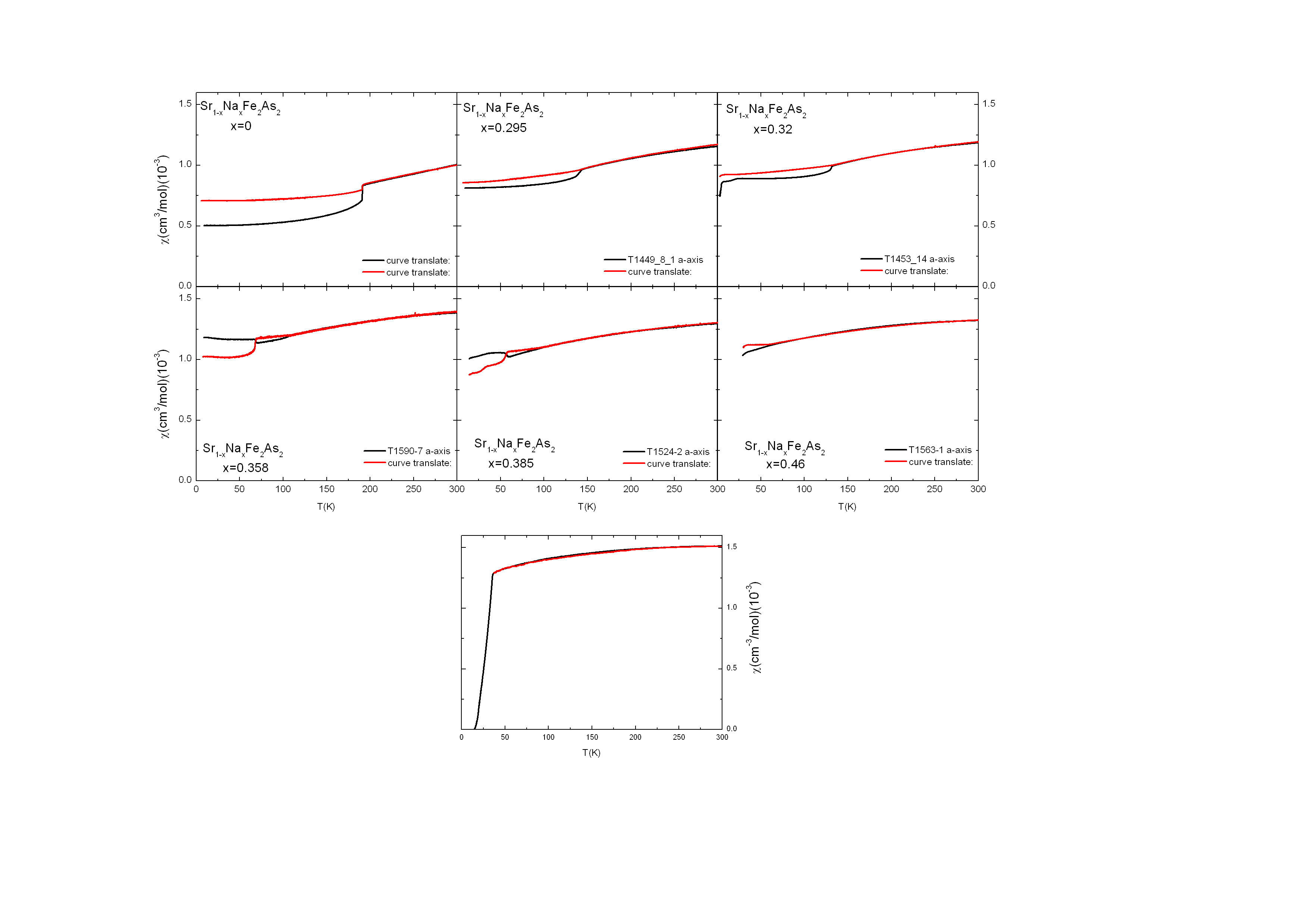}
\end{minipage}
\caption{Uniform magnetic susceptibility at 12 Tesla for fields parallel and perpendicular to the crystallographic $c$-axis after shifting the H-parallel-c data to match the $H$-parallel-$a$ data above the SDW transition.  As can now be clearly seen, the susceptibility of both sets of data have exactly the same temperature dependence above the SDW transition}
\label{fig:experimental_2}
\end{figure*}

\newpage

\section{S2. Mean-field model for double-{\bf Q} SDW order}
\label{sm:model}

As our starting point for obtaining a phenomenological description of the
magnetic susceptibility in spin-density wave SDW states with different configurations
of the ordered magnetic moments, we take a microscopic $5$-orbital Hubbard model for the $3d$ electrons of iron-based superconductor (FeSC) parent materials. The Hamiltonian can be decomposed as
\be
H = H_{0} + H_{\mathrm{int}},
\ee
where
\be
\label{eq:hopping}
H_{0} = \sum_{\sigma}\sum_{i,j}\sum_{\mu,\nu} c_{i \mu \sigma}^{\dagger}\left( t_{ij}^{\mu\nu}  - \mu_{0} \delta_{ij}\delta_{\mu\nu} \right)c_{j \nu \sigma},
\ee
describes the hybridization of electrons on the Fe square lattice. The hopping matrix elements $t_{ij}^{\mu\nu}$ are taken from Ref.~\onlinecite{ikeda2010}. Here we let $\mu,\nu \in \{d_{xz}, d_{yz}, d_{x^2-y^2}, d_{xy}, d_{3z^{2}-r^{2}}\}$ specify the $3d$-Fe orbitals and $i,j$ run over the sites of the square lattice. The filling is fixed by the chemical potential $\mu_{0}$, and the fermionic operators $ c_{i \mu \sigma}^{\dagger}$, $c_{i \mu \sigma}$ create and destroy, respectively, an electron at site $i$ in orbital $\mu$ with spin polarization $\sigma$. The electronic Hubbard-Hund interaction Hamiltonian (with normal ordering implied) can be compactly written as
\be
\label{eq:interaction}
H_{\mathrm{int}} 
& = & -\sum_{i} \sum_{\{\mu_j \},\{ \sigma_k \}} 
[U]^{\mu_{1}\sigma_{1};\mu_{2}\sigma_{2}}_{\mu_{3}\sigma_{3};\mu_{4}\sigma_{4}}
c_{i\mu_{1}\sigma_{1}}^{\dagger}
c_{i\mu_{2}\sigma_{2}}
c_{i\mu_{3}\sigma_{3}}^{\dagger}
c_{i\mu_{4}\sigma_{4}}.
\ee
The bare interaction vertex $[U]^{\mu_{1}\sigma_{1};\mu_{2}\sigma_{2}}_{\mu_{3}\sigma_{3};\mu_{4}\sigma_{4}}$ defined above can be decomposed into charge and spin vertices in the particle-hole channel as
\be
[U]^{\mu_{1}\sigma_{1};\mu_{2}\sigma_{2}}_{\mu_{3}\sigma_{3};\mu_{4}\sigma_{4}}
= - \frac{1}{2} 
\left( 
[U_{\mathrm{c}}]^{\mu_{1}\mu_{2}}_{\mu_{3}\mu_{4}} \,
\delta_{\sigma_{1}\sigma_{2}}\delta_{\sigma_{3}\sigma_{4}}  - 
[U_{\mathrm{s}}]^{\mu_{1}\mu_{2}}_{\mu_{3}\mu_{4}} \,
{\boldsymbol \sigma}_{\sigma_{1}\sigma_{2}} \cdot {\boldsymbol \sigma}_{\sigma_{3}\sigma_{4}}
\right),
\ee
which in turn are defined by
\be
[U_{\mathrm{s}}]^{\mu\mu}_{\mu\mu} = U,
\quad
[U_{\mathrm{s}}]^{\nu\mu}_{\mu\nu} = U^{\prime},
\quad
[U_{\mathrm{s}}]^{\nu\nu}_{\mu\mu} = J,
\quad
[U_{\mathrm{s}}]^{\mu\nu}_{\mu\nu} = J^{\prime},
\quad\text{with}\,\mu \neq \nu,
\ee
and
\be
[U_{\mathrm{c}}]^{\mu\mu}_{\mu\mu} = U,
\quad
[U_{\mathrm{c}}]^{\nu\mu}_{\mu\nu} = 2J - U^{\prime},
\quad
[U_{\mathrm{c}}]^{\nu\nu}_{\mu\mu} = 2U^{\prime}-J,
\quad
[U_{\mathrm{c}}]^{\mu\nu}_{\mu\nu} = J^{\prime},
\quad\text{with}\,\mu \neq \nu,
\ee
and zero otherwise. The onsite interaction is parametrized by an intraorbital Hubbard-$U$, an interorbital coupling $U^{\prime}$, Hund's coupling $J$ and pair hopping $J^{\prime}$. We will restrict ourselves to rotationally symmetric interaction parameters, which are realized for $U^{\prime} = U - 2J$, $J = J^{\prime}$. 

Going to a Bloch-representation, neglecting the charge channel and performing a mean-field decoupling of the electronic interaction for SDW mean-fields with ordering vectors ${\bf Q}_{1} = (\pi,0)$ and ${\bf Q}_{2} = (0,\pi)$, we arrive at the mean-field Hamiltonian 
\be 
H_{\mathrm{2Q-SDW}}
=
\sum_{{\bf k}}^{\prime}
\sum_{\mu,\nu}
\Psi_{{\bf k}\mu}^{\dagger}
\begin{pmatrix}
\xi^{\mu\nu}({\bf k}) {\bf 1} & {\boldsymbol W}_{1}^{\mu\nu} \cdot {\boldsymbol \sigma} & {\boldsymbol W}_{2}^{\mu\nu} \cdot {\boldsymbol \sigma} & 0 \\
{\boldsymbol W}_{1}^{\mu\nu} \cdot {\boldsymbol \sigma} & \xi^{\mu\nu}({\bf k} + {\bf Q}_{1}) {\bf 1} & 0 & {\boldsymbol W}_{2}^{\mu\nu} \cdot {\boldsymbol \sigma} \\
{\boldsymbol W}_{2}^{\mu\nu} \cdot {\boldsymbol \sigma} & 0 & \xi^{\mu\nu}({\bf k} + {\bf Q}_{2}) {\bf 1} & {\boldsymbol W}_{1}^{\mu\nu} \cdot {\boldsymbol \sigma} \\
0 & {\boldsymbol W}_{2}^{\mu\nu} \cdot {\boldsymbol \sigma} & {\boldsymbol W}_{1}^{\mu\nu} \cdot {\boldsymbol \sigma} & \xi^{\mu\nu}({\bf k} + {\bf Q}_{3}) {\bf 1}
\end{pmatrix}
\Psi_{{\bf k}\nu},
\ee 
where the ${\bf k}$-sum runs over momenta in the reduced Brillouin zone (rBZ), $ {\bf k} \in [-\pi/2,\pi/2) \times [-\pi/2,\pi/2) $, and $\xi^{\mu\nu}({\bf k}) = \epsilon^{\mu\nu}({\bf k}) - \mu_{0} \delta_{\mu\nu}$, where $\epsilon^{\mu\nu}({\bf k})$ denotes the Fourier transformed hopping matrix. In order to write the downfolded Hamiltonian in a compact matrix form, we introduced the spinor (with ${\bf Q}_{3} = {\bf Q}_{1}+ {\bf Q}_{2} = (\pi,\pi)$)
\be 
\Psi_{{\bf k}\mu}^{\dagger} =
\begin{pmatrix}
c_{{\bf k}\mu\uparrow}^{\dagger},
c_{{\bf k}\mu\downarrow}^{\dagger},
c_{{\bf k} + {\bf Q}_{1}\mu\uparrow}^{\dagger}
c_{{\bf k} + {\bf Q}_{1}\mu\downarrow}^{\dagger}
c_{{\bf k} + {\bf Q}_{2}\mu\uparrow}^{\dagger}
c_{{\bf k} + {\bf Q}_{2}\mu\downarrow}^{\dagger}
c_{{\bf k} + {\bf Q}_{3}\mu\uparrow}^{\dagger}
c_{{\bf k} + {\bf Q}_{3}\mu\downarrow}^{\dagger}
\end{pmatrix},
\ee
and $\Psi_{{\bf k}\mu} = (\Psi_{{\bf k}\mu}^{\dagger})^{\dagger}$. We have defined the SDW mean-fields
\be 
{\boldsymbol W}_{1}^{\mu\nu} = 
-\frac{1}{2}\sum_{\mu^{\prime},\nu^{\prime}} 
[U_{\mathrm{s}}]^{\mu\nu}_{\mu^{\prime}\nu^{\prime}}
{\boldsymbol M}_{1}^{\mu^{\prime}\nu^{\prime}}, \quad
{\boldsymbol W}_{2}^{\mu\nu} = 
-\frac{1}{2}\sum_{\mu^{\prime},\nu^{\prime}} 
[U_{\mathrm{s}}]^{\mu\nu}_{\mu^{\prime}\nu^{\prime}}
{\boldsymbol M}_{2}^{\mu^{\prime}\nu^{\prime}},
\ee
with the expectation values
\be 
{\boldsymbol M}_{1}^{\mu\nu} =
\frac{1}{\mathcal{N}} \sum_{{\bf k}}\sum_{\sigma,\sigma^{\prime}}
\langle 
c_{{\bf k} + {\bf Q}_{1}\mu\sigma}^{\dagger}
{\boldsymbol \sigma}_{\sigma\sigma^{\prime}}
c_{{\bf k}\nu\sigma^{\prime}}
\rangle
, \quad
{\boldsymbol M}_{2}^{\mu\nu} =
\frac{1}{\mathcal{N}} \sum_{{\bf k}}\sum_{\sigma,\sigma^{\prime}}
\langle 
c_{{\bf k} + {\bf Q}_{2}\mu\sigma}^{\dagger}
{\boldsymbol \sigma}_{\sigma\sigma^{\prime}}
c_{{\bf k}\nu\sigma^{\prime}}
\rangle,
\ee
where $\mathcal{N}$ denotes the number of unit cells of the Fe square lattice.
The mean fields couple the components of the spinor $\Psi_{{\bf k}\mu}$ in an orbital and spin-dependent way, depending on the orientation of the magnetic moment associated with the SDW order. Here, we will not pursue a self-consistent approach, that corresponds to minimizing the mean-field free energy over the magnetic order parameters. Instead, we will parameterize the mean-fields guided by numerical solutions of the self-consistent mean-field equations studied in Refs.~\onlinecite{gastiasoro2015,scherer2016}. Since we neglect spin-orbit coupling, all collinear SDW configurations related by a spin-rotation are energetically equivalent. While we have no predictive power concerning the microscopic mechanism determining the actual spin orientation, we use this freedom to choose a moment orientation that produces a magnetic anisotropy in the uniform susceptibility, that is compatible with the experimental result.

\section{S2. Mean-field parameters}
\label{sm:parameters}

Here, we briefly describe the parameterization of the magnetic states
in the framework of our phenomenological mean-field model for a metallic state with
single- or double-Q antiferromagnetic SDW order. The mean fields ${\boldsymbol W}_{1}^{\mu\nu}$ and ${\boldsymbol W}_{2}^{\mu\nu}$ are specified by a choice for the interaction parameters $U$ and $J$ entering the spin-vertex
$ [U_{\mathrm{s}}]^{\mu_{1}\mu_{2}}_{\mu_{3}\mu_{4}} $ and the order-parameters
$ {\boldsymbol M}_{1}^{\mu\nu} $ and $ {\boldsymbol M}_{2}^{\mu\nu} $. Motivated by phase diagrams obtained from self-consistent solutions of the mean-field equations~\cite{gastiasoro2015,scherer2016}, we fix the Hubbard-$U$ to $ U = 0.95\,$eV and $ J = U/4 $. The aforementioned numerical results further indicate, that inter-orbital $\mu \neq \nu$ elements typically play a subleading role and will hence be neglected in the following. We therefore take
\be 
{\boldsymbol M}_{1}^{\mu\nu} = \alpha_{1} \delta_{\mu\nu} \left( m_{1,\mu}^{x}, m_{1,\mu}^{y},  m_{1,\mu}^{z}\right)^{T} , \quad
{\boldsymbol M}_{2}^{\mu\nu} = \alpha_{2} \delta_{\mu\nu} \left( m_{2,\mu}^{x}, m_{2,\mu}^{y},  m_{2,\mu}^{z}\right)^{T}.
\ee
We note that by virtue of the Kronecker delta $\delta_{\mu\nu}$ in orbital space, the interorbital repulsion $U^{\prime}$ and the pair-hopping interaction $J^{\prime}$ do not contribute to the mean-field self energy. As we are basically interested in interpolating between a stripe configuration with $ m_{1,\mu}^{i} \neq 0 $, $ m_{2,\mu}^{\mu} = 0 $ and a $ C_{4} $ symmetric CSDW state, we proceed by fixing the elements $ m_{1,\mu}^{i} $, and subsequently generate $ m_{2,\mu}^{i}  $ by applying a $C_{4}$ transformation in orbital space to $ {\boldsymbol M}_{1}^{\mu\nu} $, while assuming a collinear orientation. The parameters $\alpha_{1}$ and $\alpha_{2}$ fix the size of the magnetic moments ${\boldsymbol m}_{1}$ and ${\boldsymbol m}_{2}$ associated with the spin-density waves with ordering vectors ${\bf Q}_{1}$ and ${\bf Q}_{2}$, respectively. We note that $ |{\boldsymbol m}_{1/2} | = | \sum_{\mu}{\boldsymbol M}_{1/2}^{\mu\mu} |  = \alpha_{1/2} $ is achieved by a corresponding normalization of the $ m_{1/2,\mu}^{i} $. Below, we collect the details of the magnetic configurations entering the simulation of the magnetic susceptibility.

\bigskip

\noindent {\bf Single-Q stripe SDW state with in-plane moment}

\noindent To model a stripe SDW with ordering vector ${\bf Q}_{1}$ and moment orientation along the $x$-axis ($a$ direction of the Fe square lattice), we take
\be
m_{1,xz}^{x} = 0.1481, \quad
m_{1,yz}^{x} = 0.3333, \quad
m_{1,x^{2}-y^{2}}^{x} = 0.0741, \quad
m_{1,xy}^{x} = 0.2963, \quad
m_{1,3z^{2}-r^{2}}^{x} = 0.1481,
\ee
satisfying $ \sum_{\mu}  m_{1,\mu}^{x} = 1$, and $ m_{2,\mu}^{y} = m_{2,\mu}^{z} = 0$ for all $\mu$. The above choice reflects the relative orbital contributions found in numerical solutions of the self-consistent mean-field equations. We thus obtain
\be 
{\boldsymbol M}_{1}^{\mu\nu} = \alpha_{1} \delta_{\mu\nu} m_{1,\mu}^{x} \left( 1, 0,  0\right)^{T},
\ee
with moment-size adjustable by the parameter $ \alpha_{1} $. 

\bigskip

\noindent {\bf Single-Q stripe SDW state with out-of-plane component}

\noindent An out-of-plane component along the $z$-axis is now easily generated by applying a rotation about the $y$-axis with rotation angle $\theta$. We arrive at
\be 
{\boldsymbol M}_{1}^{\mu\nu} = \alpha_{1} \delta_{\mu\nu} m_{1,\mu}^{x} 
\left(\cos(\theta), 0,  \sin(\theta)\right)^{T}.
\ee

\bigskip

\noindent{\bf Double-Q CSDW state with non-equal moments}

\noindent Finally, we let the parameter $ \alpha_{2} $ be non-zero in order to include a second SDW component with ordering vector ${\bf Q}_{2}$ with a moment oriented parallel to the moment of the first SDW. Noting that the $m_{2,\mu}^{x}$ are generated by a $C_{4}$ transformation from  $m_{1,\mu}^{x}$, we thus simply arrive at
\be 
{\boldsymbol M}_{1}^{\mu\nu} = \alpha_{1} \delta_{\mu\nu} m_{1,\mu}^{x} 
\left(\cos(\theta), 0,  \sin(\theta)\right)^{T}, \quad
{\boldsymbol M}_{2}^{\mu\nu} = \alpha_{2} \delta_{\mu\nu} m_{2,\mu}^{x} 
\left(\cos(\theta), 0,  \sin(\theta)\right)^{T}.
\ee 
We note, that the magnetic state is $C_{4}$ symmetric only in the case $\alpha_{1} = \alpha_{2}$. Deviations from $C_{4}$ symmetry will manifest in e.g. the reconstruction of the electronic bandstructure of the paramagnetic state by the finite SDW order.

\section{S3. Uniform susceptibility in presence of double-${\bf Q}$ SDW order}
\label{sm:phbub}

To access the magnetic susceptibility, we define the connected, time-ordered (with $\mathcal{T}_{\tau}$ the imaginary-time ordering operator) Matsubara spin-spin correlation function
\be
\chi^{ij}(\mathrm{i}\omega_n,{\bf q}) = g^2\int_{0}^{\beta} \! d\tau \,
\mathrm{e}^{\mathrm{i}\omega_n \tau}
\langle \mathcal{T}_{\tau} S^{i}_{{\bf q}}(\tau) S^{j}_{-{\bf q}}(0)\rangle_{c},
\ee
with the $i$th component ($ i = x,y,z $) of the Fourier transformed electron-spin operator for the 1-Fe unit cell given as (with $\sigma^{i}$ denoting Pauli matrices)
\be
S^{i}_{{\bf q}}(\tau) = \frac{1}{\sqrt{\mathcal{N}}}\sum_{{\bf k},\mu,\sigma,\sigma^{\prime}} c_{{\bf k} - {\bf q}\mu\sigma}^{\dagger}(\tau) \frac{\sigma_{\sigma\sigma^{\prime}}^{i}}{2}c_{{\bf k}\mu\sigma^{\prime}}(\tau).
\ee
In the following, we will approximate the static uniform susceptibility in the
SDW-ordered state by the summed particle-hole bubble, where the electronic propagator includes the mean-field self-energy introduced above.
To this end, we define a generalized Matsubara correlation function
\be
\label{eq:genchi}
[\chi]^{\mu_1\sigma_1; \mu_2\sigma_2}_{\mu_3\sigma_3; \mu_4\sigma_4}(\mathrm{i}\omega_n,{\bf q})=
\frac{1}{\beta\mathcal{N}}\int_{0}^{\beta} \! d\tau \, \mathrm{e}^{\mathrm{i}\omega_n \tau}\sum_{{\bf k},{\bf k}^{\prime}}
\langle \mathcal{T}_{\tau} 
c_{{\bf k} - {\bf q}\mu_1\sigma_1}^{\dagger}(\tau) 
c_{{\bf k}\mu_2\sigma_2}(\tau)
c_{{\bf k}^{\prime} + {\bf q}\mu_3\sigma_3}^{\dagger}(0) 
c_{{\bf k}^{\prime}\mu_4\sigma_4}(0)
\rangle_{c}.
\ee
Applying Wick's Theorem and neglecting interaction effects beyond the mean-field self-energy, the correlation function turns into a summed particle-hole bubble with dressed electronic propagators. Evaluating the ensuing Matsubara sum over fermionic Matsubara frequencies, we obtain
\be 
[\chi]^{\mu_1\sigma_1;\mu_2\sigma_2}_{\mu_3\sigma_3;\mu_4\sigma_4}(\mathrm{i}\omega_n,{\bf q}) & = & -\frac{1}{\mathcal{N}}\sum_{{\bf k}}^{\prime}\sum_{n_1,n_2}
[\mathcal{M}_{n_1,n_2}({\bf k},{\bf q})]^{\mu_1\sigma_1;\mu_2\sigma_2}_{\mu_3\sigma_3;\mu_4\sigma_4}
\frac{f(\epsilon_{n_1}({\bf k}-{\bf q})) - f(\epsilon_{n_2}({\bf k}))}{\mathrm{i}\omega_n + \epsilon_{n_1}({\bf k}-{\bf q}) - \epsilon_{n_2}({\bf k})},
\ee
with the eigenenergies $\epsilon_{n}(\bf k)$ of the Hartree-Fock Hamiltonian and $f(\epsilon)$ the Fermi-Dirac distribution. The orbital-to-band matrix elements entering the components of the correlation function are collected in the pre-factor $[\mathcal{M}_{n_1,n_2}({\bf k},{\bf q})]^{\mu_1\sigma_1;\mu_2\sigma_2}_{\mu_3\sigma_3;\mu_4\sigma_4}$, that reads
\be 
[\mathcal{M}_{n_1,n_2}({\bf k},{\bf q})]^{\mu_1\sigma_1;\mu_2\sigma_2}_{\mu_3\sigma_3;\mu_4\sigma_4}=
\sum_{\{l_1,l_2,l_3,l_4\}}
\mathcal{U}_{{\mu_1l_1\sigma_1,n_1}}^{\ast}({\bf k} - {\bf q})
\mathcal{U}_{{\mu_2l_2\sigma_2,n_2}}({\bf k})
\mathcal{U}_{{\mu_3l_3\sigma_3,n_2}}^{\ast}({\bf k})
\mathcal{U}_{{\mu_4l_4\sigma_4,n_1}}({\bf k} - {\bf q}).
\ee
Here, $\sum_{\{l_1,l_2,l_3,l_4\} }\cdots $ denotes a restricted sum over $l$-index tuples contributing to correlation function and the prime on the sum denotes a ${\bf k}$-summation over the corresponding reduced Brillouin zone. The unitary matrix $\mathcal{U}_{{\mu l\sigma,n}}({\bf k})$ diagonalizes the mean-field Hamiltonian. The spinor $\Psi_{{\bf k}\mu l\sigma}$ (where now we have made all quantum numbers explicit) transforms as $\Psi_{{\bf k}\mu l\sigma} = \sum_{n}\mathcal{U}_{{\mu l\sigma,n}}({\bf k}) \Phi_{{\bf k}n}$. Performing analytic continuation $\mathrm{i}\omega_n \to \omega + \mathrm{i}\eta$ ($\eta \to 0^{+}$) and taking the uniform limit of the static susceptibility, we obtain the susceptibility tensor
\be 
\chi^{ij} = \lim_{{\bf q}\to{\bf 0},\omega\to 0}
\frac{g^{2}}{4} 
\sum_{\sigma_{1},\sigma_{2},\sigma_{3},\sigma_{4}}
\sigma^{i}_{\sigma_{1}\sigma_{2}} \sigma^{j}_{\sigma_{3}\sigma_{4}}
\sum_{\mu,\nu}
[\chi]^{\mu\sigma_1;\mu\sigma_2}_{\nu\sigma_3;\nu\sigma_4}(\omega,{\bf q}).
\ee
In order to separate out the anisotropic contributions, we decompose the diagonal elements of the susceptibility tensor as $\chi^{ii} = \chi_{0} + \Delta\chi^{ii}$,
where the isotropic contribution is given by
\be 
\chi_{0} = \frac{1}{4}\lim_{{\bf q}\to{\bf 0},\omega\to 0}
\frac{g^{2}}{4} 
\sum_{\sigma,\sigma^{\prime}}
\sum_{\mu,\nu}
[\chi]^{\mu\sigma;\mu\sigma^{\prime}}_{\nu\sigma^{\prime};\nu\sigma}(\omega,{\bf q}).
\ee
%


%

\end{widetext}

\end{document}